\documentclass[aps,twocolumn,showpacs,preprintnumbers,amsmath,amssymb,superscriptaddress]{revtex4}

\usepackage{bm}

\newcommand{\rmd}{{\mathrm{d}}}
\newcommand{\rme}{{\mathrm{e}}}
\newcommand{\rmi}{{\mathrm{i}}}

\begin{document}

\title {Self-consistency vanishes in the plateau regime of the bump-on-tail instability}

\author{D.F.~Escande, Y.~Elskens}
\email{dominique.escande@univ-provence.fr,yves.elskens@univ-provence.fr}

\affiliation{UMR 6633 CNRS--Universit\'{e} de Provence, Marseille, France}

\date{\today}

\begin{abstract}
Using the Vlasov-wave formalism, it is shown that self-consistency vanishes in the plateau regime of the bump-on-tail instability if the plateau is broad enough. This shows that, in contrast with the ``turbulent trapping" Ansatz, a renormalization of the Landau growth rate or of the quasilinear diffusion coefficient is not necessarily related to the limit where the Landau growth time becomes large with respect to the time of spreading of the particle positions due to velocity diffusion.
\end{abstract}

%
%

\maketitle

\section{Introduction}

The saturation of the bump-on-tail instability is a tough problem of kinetic plasma physics, which is still the subject of a controversy \cite{LP,livre}. It was originally tackled in the frame of the Vlasov-Poisson formalism through the quasilinear approximation that neglects mode coupling \cite{DP,VV}. However mode coupling was proved to be important during the saturation of the instability \cite{ALP}, which called for a new calculation scheme tailored to this phase. This is generically difficult because the particle motion is chaotic ; moreover wave-particle self-consistency makes the description of this chaotic regime especially difficult. This paper shows that, if the tail particle distribution displays a plateau, the waves whose phase velocity belongs to the plateau feel a negligible mode-coupling in the limit where the plateau is broad enough.

\section{Description of wave-particle self-consistency}

The difficulty to describe the nonlinear regime of the Vlasov-Poisson system of equations, and the progress in the chaotic dynamics of Hamiltonian systems with a finite number of degrees of freedom were an incentive to tackle the description of the saturation regime with the so-called self-consistent Hamiltonian that describes the one-dimensional self-consistent evolution, in a plasma with spatial periodicity $L$, of $M$ Langmuir waves with $N$ particles in the tail of the electron distribution function per length $L$ \cite{livre}. In the absence of tail particles, the Langmuir waves are the collective motions of a plasma without resonant particles with a density $n_{\rm bulk}$, and with a plasma frequency $\omega_{\rm p}$. Wave $j$ has pulsation $\omega_{j}$ related to the wavenumber $k_j$ (a multiple of $2 \pi /L$) through the Bohm-Gross dispersion relation $\epsilon(k_j,\omega_j)=0$ for the plasma dielectric function $\epsilon$. Each wave may be viewed as a harmonic oscillator. The interaction of these waves with the $N$ tail particles is described by the self-consistent Hamiltonian
\begin{eqnarray}
  H_{\rm sc} =& & \sum_{r=1}^N {\frac {p_r^2} {2}}
        + \sum_{j=1}^M \omega_{j0} {\frac {X_j^2 + Y_j^2} 2}
        \nonumber\\
        &+& \varepsilon \sum_{r=1}^N \sum_{j=1}^M
           k_j^{-1} \beta_j (Y_j \sin k_j x_r - X_j \cos k_j x_r),
   \label{eqHXY}
\end{eqnarray}
where $(x_r,p_r)$ are the conjugate position and momentum of particle $r$ (with mass normalized to unity), $(X_j,Y_j)$ are the conjugate generalized coordinate and momentum of the harmonic oscillator corresponding to Langmuir wave number $j$, $\beta_j = [\partial \epsilon / \partial \omega (k_j,\omega_j) ]^{-1/2}$, and
\begin{equation}
  \varepsilon
  =
  \omega_{\rm p} \sqrt{ \frac {2 m n_{\rm tail}} {n_{\rm bulk}N}}
\label{eps}
\end{equation}
is the coupling coefficient, where $m$ is the electron mass and $n_{\rm tail}$ is the density of the tail particles.
This Hamiltonian generates the evolution equations
\begin{eqnarray}
  \dot x_r
  &=&
  p_r
  \label{eqxx}
  \label{SZ110}
    \\
  \dot p_r
  &=&
  \varepsilon
  \Re \Bigl( \sum_{j=1}^M \rmi  \beta_j Z_j \rme^{\rmi  k_j x_r}
      \Bigr)
  \label{eqpp}
    \\
  \dot Z_j
  &=&
  - \rmi  \omega_{j} Z_j
  + \rmi  \varepsilon \beta_j k_j^{-1} \sum_{r=1}^N
                 \rme^{- \rmi  k_j x_r}
  \label{eqZ}
\end{eqnarray}
where $\Re$ is the real part. Equation (\ref{eqpp}) makes clear the link between $Z_j = X_j  + \rmi Y_j $ and the electric field of wave $j$.

For the present paper, it is easier to describe the tail particles through a velocity distribution function $f(x,p,t)$. This is possible through the so-called Vlasov-wave model that is obtained as a mean-field limit (limit $N \to \infty$) of the dynamics defined by $H_{\rm sc}$ \cite{FE,EF}, in analogy to that performed by Spohn for his elegant and short derivation of the Vlasov equation \cite{Sp}. This yields the following set of coupled equations
\begin{equation}
  \partial_t f
  + p \partial_x f
  + \Re \Bigl( \sum_{j=1}^M \zeta_j \rme^{\rmi  (k_j x - \omega_j t) }
        \Bigr)
        \partial_p f
  = 0,
\label{Vlasov2}
\end{equation}
\begin{equation}
  \dot \zeta_j
  =
  -  \varepsilon_j
    \int_{\Lambda}
      \rme^{- \rmi  (k_j x - \omega_j t)} f(p,x,t) \rmd p \rmd x ,
\label{eqZV2}
\end{equation}
where $\Lambda = [0,L] \times {\mathbb R}$ is the one-particle $(x,p)$ space,
and with the rescaled wave complex envelope variable
\begin{equation}
  \zeta_j
  =
  \rmi \varepsilon \beta_j Z_j \rme^{\rmi \omega_j t}
\label{defzeta}
\end{equation}
and coupling constant
\begin{equation}
  \varepsilon_j
  =
  2 k_j^{-1} \omega_{\rm p}^2 \beta_j^2 m \frac {n_{\rm tail}} {n_{\rm bulk}}
\label{varepsilonprime}  \, .
\end{equation}
Here $\int_{\Lambda} f(p,x,t) \rmd p \rmd x =1$. This model keeps some symmetry in the description of waves and particles, but gets rid of the granularity effects due to the discrete description of particles, like spontaneous emission of waves by particles.

\section{Dynamics when the distribution is a plateau}

This paper considers the dynamics defined by Eqs (\ref{Vlasov2}-\ref{eqZV2}), while starting at time $t=0$ from (i)~a spectrum of Langmuir waves where all nearby waves are in resonance overlap and (ii)~a spatially uniform plateau for the particle velocity distribution function over this overlap domain with a height $f_0$. We first start with a simplistic description of this dynamics, which will be useful to derive a more accurate one hereafter: since the plateau is spatially uniform, there is no source term for the waves in Eq. (\ref{eqZV2}), which keep constant complex amplitudes; therefore the particle dynamics is that defined by a prescribed spectrum of waves, which preserves the initial plateau. Clumps of particles may experience a strong turbulent trapping, but the distribution function is unaffected by this granular effect.

In reality Kolmogorov-Arnold-Moser (KAM) tori bounding the chaotic domain defined by a prescribed spectrum of waves experience a sloshing motion due to the waves. This brings a small spatial modulation to the particle density which provides a source term for the Langmuir waves in Eq. (\ref{eqZV2}). However the evolution of the wave spectrum is slow, which only brings a small change to the previous simplistic picture. This slow evolution suggests to introduce an adiabatic description of the true dynamics.

To be specific, we consider the case where the wave phase velocities range over an interval $[v_0, v_1]$, the nearest KAM tori have velocities $v_{\mathrm a}$ and $v_{\mathrm b}$ and the particle distribution function plateau ranges over $[u_0,u_1]$, with $u_0 \lesssim v_{\mathrm a} \lesssim v_0 < v_1 \lesssim v_{\mathrm b} \lesssim u_1$. Therefore the plateau width $\Delta v = u_1 - u_0$ is essentially equal to the chaotic domain width $v_{\mathrm b} - v_{\mathrm a}$ and to the wave spectrum width $u_1 - u_0$. Moreover, the wave phase numbers are all of comparable order of magnitude, typically $k$. Particles with velocity $v$ experience the oscillations of a wave $j$ at relative frequencies $\Omega_j = \omega_j - k_j v$, and for a particle or a wave with velocity $v$ the nearest KAM velocity defines $\Omega_{\min} = k \min(v_{\mathrm b} - v, v - v_{\mathrm a})$. The resonance overlap in the wave spectrum is characterized by the small parameter $k^{1/2} \delta \Omega / \zeta_*^{1/2} \ll 1$, where $\delta \Omega$ is the (Doppler) frequency detuning of two nearby waves and $\zeta_*$ is the typical modulus of a wave amplitude.

We now formalize this adiabatic description. Let $\zeta_j(t)$ be the value of $\zeta_j$ at time $t$ in the self-consistent dynamics. Consider the non self-consistent dynamics ${\rm D} (t_0)$ defined by the self-consistent spectrum of Langmuir waves frozen at time $t_0$, as defined by Eq. (\ref{Vlasov2}) where the $\zeta_j$'s are substituted with the $\zeta_j (t_0)$'s. The chaotic domain ${\rm C} (t_0)$ in single-particle (Boltzmann or $\mu$) phase space $(x,p)$ defined by this frozen wave spectrum is bounded above and below in $p$ by two KAM tori, respectively ${\rm T_a} (t_0)$ and ${\rm T_b} (t_0)$. The initial particle distribution function $f(x,p,0)$ is assumed to be uniform on ${\rm C} (0)$ ; let $f_0$ be this uniform value. During the adiabatic evolution corresponding to the true self-consistent dynamics, $f(x,p,t)$ stays uniform on ${\rm C} (t)$ and keeps the value $f_0$. We are left with the calculation of the modulation of the width of a single water bag with height $f_0$. The modulation of this width is given by that of KAM tori ${\rm T_a} (t_0)$ and ${\rm T_b} (t_0)$, which may be computed by perturbation theory in the typical amplitude $\zeta_*$ of the Langmuir waves.

Since mode-mode coupling is a four-wave process, the first non vanishing contribution to Eq. (\ref{eqZV2}) is of order $\zeta_*^3$. For waves with phase velocities further than $\Omega_{\min}/k$ from the edges of the plateau, a pessimistic estimate for $\dot \zeta_j (t)$ turns out to scale like $\zeta_* \Omega_{\min} / (\Delta v \delta \Omega)$. With the (also pessimistic) estimate $\Omega_{\min} \propto \zeta_*^{1/2}$ associated with the modulation of KAM tori, $\dot \zeta_*$ scales like $\zeta_*^{3/2} / (\Delta v \delta \Omega)$, for a given spectrum with fixed discretization $\delta \Omega$, and thus vanishes if the plateau is broad enough ($\Delta v \to \infty$). This justifies a posteriori our previous adiabatic approximation. As a result the plateau dynamics (further than $\Omega_{\min}/k$ from its boundaries) is almost the same as in a prescribed field of Langmuir waves. Therefore the chaotic motion of particles is almost unchanged due to the nonlinear coupling of Langmuir waves.

\section{Conclusion}

We have just shown that self-consistency vanishes in the plateau regime of the bump-on-tail instability if the plateau is broad enough. This means that the diffusion coefficient $D(p)$ of particles with momentum $p$ is that found for the dynamics of particles in a prescribed spectrum of Langmuir waves. Let $D_{\rm QL}(p)$ be the quasilinear value of this coefficient. In the resonance overlap regime $D/D_{\rm QL}$ may cover a large range of values \cite{CEV,livre,YE}. In particular $D \simeq D_{\rm QL}$ is obtained for random phases of the waves and strong resonance overlap \cite{CEV,livre,YE}.

Let $\gamma_{\rm Landau}$ be the maximum Landau growth rate of Langmuir waves for the instantaneous value of $f$. Let $ \tau_{\rm spread} = (k^2 D)^{-1/3}$ be the time after which the ballistic approximation fails for particles diffusing with the maximum instantaneous $D$ due to a spectrum of Langmuir waves with typical wavenumber $k$. The plateau regime corresponds to $\gamma_{\rm Landau} \tau_{\rm spread} = 0$. Since $D/D_{\rm QL}$ may cover a large range of values in this regime, $\gamma_{\rm Landau} \tau_{\rm spread} \ll 1$ does not imply any renormalization of $D/D_{\rm QL}$, nor of $\gamma / \gamma_{\rm Landau}$ by wave-particle momentum conservation, in contrast with the ``turbulent trapping" Ansatz of \cite{LPTT}. The value of $D/D_{\rm QL}$ in the plateau regime of the bump-on-tail instability depends on the kind of wave spectrum the beam-plasma system reaches during the saturation phase of the instability. A numerical simulation using the Vlasov-wave model is now attempting to uncover this spectrum.


\begin{thebibliography}{33}

\bibitem{LP} Laval G and Pesme D 1999
  Controversies about quasilinear theory
  {\it Plasma Phys. Control. Fusion} {\bf 41} A239--A246

\bibitem{livre}
Elskens Y and Escande D F 2003
  \textit{Microscopic Dynamics of Plasmas and Chaos}
  (Bristol: Institute of Physics Publishing)

\bibitem{VV} Vedenov A A, Velikhov E P and Sagdeev R Z 1962
  Quasilinear theory of plasma oscillations
  {\it Nuclear Fusion Suppl.} {\bf 2} 465--475

\bibitem{DP} Drummond W E and Pines D 1962
  Nonlinear stability of plasma oscillations
  {\it Nuclear Fusion Suppl.} {\bf 3} 1049--1057

\bibitem{ALP} Adam J C, Laval G and Pesme D 1979
  Reconsideration of quasilinear theory
  {\it Phys. Rev. Lett.} {\bf 43} 1671--1675

\bibitem{FE} Firpo M-C and Elskens Y 1998
  Kinetic limit of $N$-body description of wave-particle
  self-consistent interaction
  {\it J. Stat. Phys.} {\bf 93} 193--209

\bibitem{EF} Elskens Y and Firpo M-C 1998
  Kinetic theory and large-$N$ limit for wave-particle
  self-consistent interaction
  {\it Physica Scripta} {\bf T 75} 169--172

\bibitem{Sp} Spohn H 1991
  {\it Large scale dynamics of interacting particles}
  (Berlin: Springer)

\bibitem{CEV} Cary J R, Escande D F and Verga A D 1990
  Non quasilinear diffusion far from the chaotic threshold
  {\it Phys. Rev. Lett.} {\bf 65} 3132--3135

\bibitem{YE} Elskens Y 2008
  Nonquasilinear evolution of particle velocity in incoherent waves
  with random amplitudes
  {\it Commun. Nonlinear Sci. Numer. Simul.} (in press)
  arXiv : 0803.3962 [nlin.CD]

\bibitem{LPTT} Laval G and Pesme D 1984
  Self-consistency effects in quasilinear theory~: a model for
  turbulent trapping
  {\it Phys. Rev. Lett.} {\bf 53} 270--273

\end{thebibliography}
\end{document}